\documentclass[apjl]{emulateapj}
\RequirePackage{lineno}
\usepackage{amsmath}
\usepackage{amssymb}
\usepackage{amsthm}
\usepackage{natbib,aasdefs,url,bm}
\usepackage{array}
\usepackage{float}
\usepackage{graphicx}
\usepackage{subfigure}
\usepackage{color}
\usepackage{hyperref}
\newcounter{ichi}
\newcounter{ni}
\newcounter{san}
\newcounter{yon}
\def\be{\begin{equation}}
\def\ee{\end{equation}}
\def\ba{\begin{eqnarray}}
\def\ea{\end{eqnarray}}

\def\aap{\ {A\&A}\ }

\def\apjl{\ {ApJL}\ }
\def\apjs{\ {ApJS}\ }

\shorttitle{Constraints on the Blazar Contribution to the Cumulative High-Energy Neutrino Intensity}
\shortauthors{Yuan, Murase and M\'esz\'aros}

\begin{document}

\title{Complementarity of Stacking and Multiplet Constraints on the Blazar Contribution to the Cumulative High-Energy Neutrino Intensity}

\author{Chengchao Yuan\altaffilmark{1}}\email{cxy52@psu.edu}
\author{Kohta Murase\altaffilmark{1,2} }
\author{Peter M\'esz\'aros\altaffilmark{1}}  
\altaffiltext{1}{Department of Physics; Department of Astronomy \& Astrophysics; Center for Particle and Gravitational Astrophysics, The Pennsylvania State University, University Park, PA 16802, USA}
\altaffiltext{2}{Center for Gravitational Physics, Yukawa Institute for Theoretical Physics, Kyoto University, Kyoto, Kyoto 606-8502, Japan}

\begin{abstract}
We investigate the blazar contribution to the cumulative neutrino intensity assuming a generic relationship between neutrino and gamma-ray luminosities, $L_{\nu}\propto (L_{\rm ph})^{\gamma_{\rm lw}}$. Using the gamma-ray luminosity functions for blazars including flat spectrum radio quasars (FSRQs) and BL Lac objects, as well as the $Fermi$-LAT detection efficiency, we estimate contributions from {blazars resolved by $Fermi$-LAT as well as the unresolved counterpart.} Combining the existing upper limits from stacking analyses, the cumulative neutrino flux from all blazars ({including $Fermi$-LAT resolved and unresolved ones}) are constrained {  in the range $0\lesssim\gamma_{\rm lw}\lesssim2.5$}. We also evaluate the effects of the redshift evolution and the effective local number densities for each class of FSRQs, BL Lacs, and all blazars, by which we place another type of constraints on the blazar contribution using the non-detection of high-energy neutrino multiplets. We demonstrate that these two upper limits are complementary, and that the joint consideration of the stacking and multiplet analyses {   not only supports the argument that blazars are disfavored as the dominant sources of the 100-TeV neutrino background, but it extends this argument by including also $Fermi$-LAT-unresolved blazars as well, for a more generic luminosity correlation $L_{\nu}\propto (L_{\rm ph})^{\gamma_{\rm lw}}$}.
\end{abstract}

\keywords{galaxies: active -- neutrinos}
\section{Introduction}\label{sec:intro}
Since the initial detection of high-energy astrophysical neutrinos by the IceCube Neutrino Observatory \citep{aartsen2013first,icecube2013evidence}, a cumulative flux of astrophysical neutrinos in the energy range from  $\sim$ 10 TeV to several PeV has been unveiled and measured to a higher precision \citep{aartsen2014observation,aartsen2015combined,aartsen2016observation}. The isotropic distribution of the cumulative flux as well as the background-only results from recent searches for point-like sources and multi-messenger analyses support an extragalactic origin of these neutrinos \citep{ahlers2014probing,aartsen2014searches,aartsen2015searches}. Up to now, however, the main origin of the cumulative neutrinos still remains unknown. 

The flavor ratio measured at Earth, $(\nu_e:\nu_\mu:\nu_\tau)\approx (1:1:1)$, is consistent with the prediction from the long-distance oscillations of neutrinos produced through pion decays \citep{aartsen2015flavor}, which provides one common framework for the astrophysical models dedicated to explain the cumulative neutrino flux. Many candidates have been proposed and studied \citep[see][for review]{ahlers2015high,meszaros2017astrophysical}. 
% including gamma-ray bursts \citep[GRBs,][]{waxman1997high,meszaros2001tev,murase2008prompt,wang2009prompt,bustamante2015neutrino,tamborra2016inspecting}, low-luminosity GRBs \citep{murase2006high,gupta2007neutrino,murase2013tev,xiao2014neutrino,senno2016choked,denton2018bright}, radio-loud active galactic nuclei \citep[AGNs,][]{mannheim1995high,halzen1997neutrino,anchordoqui2008high,murase2014diffuse,dermer2014photopion,petropoulou2015photohadronic,blanco2017high}, star-forming/star-burst galaxies \citep{loeb2006cumulative,murase2013testing,tamborra2014star,chang2015star,senno2015extragalactic,chakraborty2015diffuse}, galaxy clusters/groups \citep{murase2008cosmic,kashiyama2014galaxy,fang2018linking,yuan2018cumulative} and some transients, e.g. tidal disruption events \citep{wang2016tidal,senno2017high,dai2017can,lunardini2017high} and blazar flares \citep{bednarek1999gamma,halzen2005high,dermer2012variable,petropoulou2016time,gao2017direct,murase2018blazar}. 
Among these candidates, blazars, which are known as a subclass of AGNs with a relativistic jet pointing nearly towards the Earth \citep{Blandford_1978,urry1995unified}, have been frequently considered as promising ultra-high-energy cosmic-ray (CR) accelerators and high-energy neutrino emitters \citep[e.g.,][]{mucke2001neutrino,Padovani:2015mba,murase2017active,2017MNRAS.468..597R}. 
%\citep{mannheim1995high,halzen1997neutrino,anchordoqui2008high,murase2014diffuse,dermer2014photopion,petropoulou2015photohadronic,blanco2017high}. 
Recently, the IceCube collaboration announced the spatial and temporal coincidence between a muon track neutrino event IceCube170922A and a blazar TXS 0506+056 \citep{icecube2018multimessenger} at the significance $\sim3\sigma$. Intuitively, if this association is physical, the intimate link between this IceCube neutrino event and the blazar may favor blazars as the main sources of the cumulative neutrino flux, but this is not the case~\citep{murase2018blazar}. 

The maximum likelihood stacking searches for cumulative neutrino flux from the second $Fermi$-LAT AGN catalog (2LAC) as well as the point-source searches using the IceCube muon track events and blazars in $Fermi$-LAT 3LAC have independently shown that {$Fermi$-LAT-resolved} blazars only contribute a small portion of the IceCube cumulative neutrino flux \citep{aartsen2017contribution,pinat2017search,hooper2018active} and the hadronic models of blazar activity are strongly constrained~\citep{neronov2017strong}, {if the specific correlation $L_{\nu}\propto L_{\rm ph}$ is assumed as a prior}. 
\cite{palladino2019interpretation} evaluated the contribution of unresolved sources, and showed that the blazar contribution to the cumulative neutrino flux is constrained unless one makes an ad hoc assumption that lower-luminosity blazars entrain a larger amount of CRs.
%Also, the non-detection of neutrinos from nearby blazars such as Mrk 421 also give interesting constraints \citep{aartsen2014searches,PhysRevD.95.103003,Aartsen2019} and may disfavor blazars as the dominant origin of IceCube neutrinos.

Here we argue that, in addition {to the stacking analysis}, the absence of clustering in high-energy neutrino events, i.e., neutrino multiplets and auto-correleation, can also provide relevant constraints on various classes of proposed sources as the dominant origin of the cumulative neutrino flux~\citep{murase2016constraining,ahlers2014pinpointing,aartsen2014searches,2017JCAP...03..057F,2017ICRC...35.1014G,2019JCAP...02..002D}. The constraints are sensitive to the redshift evolution of the sources, which are especially powerful for weakly or non-evolving sources such as BL Lac objects~\citep{murase2016constraining,murase2018blazar}. But the limits are weaker for rapidly evolving sources such as FSRQs, which could significantly alleviate the constraints, as remarked by \cite{murase2018blazar}. \cite{neronov2018self} studied the constraints on evolving blazar populations and confirmed that fast evolving sources (e.g., $\xi_z=5.0$) may indeed relax the neutrino multiplet limits. 

In this work, we consider the ``joint'' implications of these independent analyses {for the global blazar population} and extend the constraints to a common case where a generic relationship between neutrino and gamma-ray luminosities, e.g., $L_\nu\propto (L_{\rm ph})^{\gamma_{\rm lw}}$, is presumed, which is more general than what has been previously considered in such analyses. Physically, the correlation between $L_\nu$ and $L_{\rm ph}$ is determined by the interactions between particles and radiation fields inside the sources. Most of physically reasonable models developed on the basis of photohadronic (e.g., $p\gamma$) interactions predict $L_{\nu}\propto (L_{\rm ph})^{{\gamma_{\rm lw}}}$ with indices of $1.0\lesssim{\gamma_{\rm lw}}\lesssim2.0$~\citep[e.g.,][]{murase2014diffuse,2014JHEAp...3...29D,tavecchio2015high,petropoulou2015photohadronic,Padovani:2015mba,murase2016constraining,2017A&A...598A..36R,murase2018blazar,2018ApJ...854...54R}.  
The index ${\gamma_{\rm lw}}$ characterizes the source models and may deviate from this fiducial range for models with increasing complexity.  Motivated by this, we treat ${\gamma_{\rm lw}}$ as a free parameter and attempt to reveal the ${\gamma_{\rm lw}}$-dependence of the upper limits on all-blazar contributions. In addition, a new feature of our analysis is that we also consider the effect of $Fermi$-unresolved blazars. One caveat is that, in this study, we assume all sources are equal and emit steadily with a single power-law spectrum.  Prior to the the IceCube-170922A alert, IceCube collaboration has found a neutrino excess from the direction of TXS 0506+056 during a 158-day time window in 2014-2015 \citep{icecube2018neutrino}, which reveals the the transient nature of the neutrino emission. 
We need to keep in mind that the multiplet limits are stronger for flaring sources~\citep{murase2018blazar}. The stacking limits are also applicable to time-averaged emission of the flaring sources, as long as the scaling between neutrino and gamma-ray luminosities hold~\citep{murase2018blazar}.

In the first part (\S\ref{sec:all_blazar}), we calculate the ratio of neutrino fluxes from $Fermi$-LAT-resolved blazars and all blazars (including both resolved and unresolved contributions). Combining this ratio with the existing constraints on $Fermi$-LAT-resolved blazars, we estimate the upper limits for all-blazar contributions. 
The multiplet constraints are given in the second part (\S\ref{sec:multiplet}) where we also derive the effective number densities $n_0^{\rm eff}({\gamma_{\rm lw}})$ and the redshift evolution factor $\xi_z({\gamma_{\rm lw}})$ for blazars and the subclasses, FSRQs and BL Lacs. 
In either case, we use the blazar gamma-ray luminosity functions provided by \cite{ajello2015origin,ajello2012luminosity,ajello2013cosmic} to reconstruct the neutrino luminosity density. 
In \S\ref{sec:discussion} we conclude with a discussion.

\section{Implications of Stacking Limits}\label{sec:all_blazar}
Given the differential density of blazars as a function of rest-frame 100 MeV-100 GeV luminosity $L_{\rm ph}$, redshift $z$ and photon index $\Gamma$ defined by the gamma-ray flux $F\propto \varepsilon_{\rm ph}^{-\Gamma}$,
\begin{linenomath*}
\begin{equation}
\frac{d^3N_{\rm bl}}{dL_{\rm ph}dzd\Gamma}=\phi_{\rm bl}(L_{\rm ph},z)\frac{dP_{\rm bl}}{d\Gamma}\frac{dV}{dz},
\label{eq:diff_density}
\end{equation}
\end{linenomath*}
where the subscript ``bl" represents blazars considered in the calculation, $\phi_{\rm bl}(L_{\rm ph},\Gamma)={d^2N_{\rm bl}}/{dL_{\rm ph}dV}$ is the luminosity function and $dP_{\rm bl}/d\Gamma$ is the probability distribution of spectral index $\Gamma$, we can directly write down the (differential) luminosity density of neutrinos from $Fermi$-LAT-resolved blazars at redshift $z$,
\begin{linenomath*}
\begin{equation}\begin{split}
\varepsilon_\nu Q_{\varepsilon_\nu}^{({\rm bl,R})}(z,{\gamma_{\rm lw}})=&\int_{L_{\rm ph, th}}^{L_{\rm ph,max}} \int_{\Gamma_{\rm min}}^{\Gamma_{\rm max}}\mathcal C^{-1}\phi_{\rm bl}(L_{\rm ph},z) L_{\nu}(L_{\rm ph})\\
&\times\frac{dP_{\rm bl}}{d\Gamma}d\Gamma dL_{\rm ph}
\end{split}
\label{eq:neu_input}
\end{equation}
\end{linenomath*}
where $L_{\nu}\propto (L_{\rm ph})^{{\gamma_{\rm lw}}}$ is the neutrino luminosity, $L_{\rm ph,max}$ is a fixed upper limit of blazar luminosity and the lower limit $L_{\rm ph,th}(L_{\rm ph},z,\Gamma)$ is determined by the $Fermi$ LAT threshold flux $F_{100,\rm th}$ in the energy range of 100 MeV -- 100 GeV. In this equation, $\mathcal C$ is the normalization coefficient determined by $\varepsilon_{\rm CR, \max}$ and $\varepsilon_{\rm CR,\min}$, the maximum and minimum energy that CRs in blazars can achieve. Since we aim to estimate the neutrino flux from a general luminosity relationship, $L_\nu \propto (L_{\rm ph})^{\gamma_{\rm lw}}$, and the physics may be unknown for a general $\gamma_{\rm lw}$, we do not try to provide the details of the gamma-ray and neutrino radiation processes. In this work, we assume that the {maximum CR energy is the same for all blazars, as is the normalization factor once the spectral index $s$ of the IceCube neutrino flux is specified. 

Here, we present one method to rewrite the integrals in equation \ref{eq:neu_input} by incorporating the $Fermi$-LAT detection efficiency. For a blazar at redshift $z$ with the luminosity $L_{\rm ph} \propto\int_{\varepsilon_{\min}}^{\varepsilon_{\max}}F(\varepsilon)\varepsilon d\varepsilon$, where $\varepsilon_{\max}=100(1+z)\rm\ GeV$ and $\varepsilon_{\min}=100(1+z)\rm\ MeV$, and the photon index $\Gamma$, the integrated photon flux at earth can be written as
\begin{linenomath*}
\begin{eqnarray}
F_{100}(L_{\rm ph},z,\Gamma)
&=&\int_{\varepsilon_{\min}}^{\varepsilon_{\max}}F(\varepsilon) d\varepsilon \nonumber\\
&=&\frac{L_{\rm ph}}{4\pi d_L^2(z)}\times
\begin{cases}
\ln\left(\frac{\varepsilon_{\max}}{\varepsilon_{\min}}\right)\frac{1}{\varepsilon_{\max}-\varepsilon_{\min}} &\text{$\Gamma=1$}\\
\frac{\varepsilon_{\max}-\varepsilon_{\min}}{\varepsilon_{\max}\varepsilon_{\min}\ln\left(\frac{\varepsilon_{\max}}{\varepsilon_{\min}}\right)} &\text{$\Gamma=2$}\\
\frac{2-\Gamma}{1-\Gamma}\frac{\varepsilon_{\max}^{1-\Gamma}-\varepsilon_{\min}^{1-\Gamma}}{\varepsilon_{\max}^{2-\Gamma}-\varepsilon_{\min}^{2-\Gamma}}&\text{$\Gamma\neq1,2$},
\end{cases}
\label{eq:flux}
\end{eqnarray}
\end{linenomath*}
where $d_L$ is the luminosity distance between the blazar and the detector. Then the lower limit of the integral in equation \ref{eq:neu_input} can be obtained by requiring $F_{100}(L_{\rm ph,th},z,\Gamma)=F_{100,\rm th}$. Alternatively, thanks to the $Fermi$-LAT detection efficiency $\epsilon(F_{100})$ provided by \cite{abdo2010fermi}, we can simplify equation \ref{eq:neu_input} by using the equivalent detection efficiency $\epsilon(L_{\rm ph},z,\Gamma)=\epsilon(F_{100})$,
\begin{linenomath*}
\begin{equation}\begin{split}
\varepsilon_\nu Q_{\varepsilon_\nu}^{({\rm bl,R})}(z,{\gamma_{\rm lw}})=&\int_{L_{\rm ph, min}}^{L_{\rm ph,max}} \int_{\Gamma_{\rm min}}^{\Gamma_{\rm max}}\mathcal C^{-1}\phi_{\rm bl}(L_{\rm ph},z) L_{\nu}(L_{\rm ph})\\
&\times\epsilon(L_{\rm ph},z,\Gamma)\frac{dP_{\rm bl}}{d\Gamma}d\Gamma dL_{\rm ph},
\end{split}
\label{eq:res_neu_input}
\end{equation}
\end{linenomath*}
where the lower limit $L_{\rm ph,min}$ reduces to a constant and represents the minimal luminosity of blazars that are considered in this work. To eliminate the instrumental selection effect produced by the low detection efficiency for dimmer blazars and to take all blazars into account, we replace the $L_{\rm ph,th}$ in equation \ref{eq:neu_input} by $L_{\rm ph,min}$, which yields the neutrino luminosity density from all blazars $\varepsilon_\nu Q_{\varepsilon_\nu}^{({\rm bl,all})}(z,{\gamma_{\rm lw}})$, which can be written explicitly as
\begin{linenomath*}
\begin{equation}\begin{split}
\varepsilon_\nu Q_{\varepsilon_\nu}^{({\rm bl,all})}(z,{\gamma_{\rm lw}})=&\int_{L_{\rm ph, min}}^{L_{\rm ph,max}} \int_{\Gamma_{\rm min}}^{\Gamma_{\rm max}}\mathcal C^{-1}\phi_{\rm bl}(L_{\rm ph},z) L_{\nu}(L_{\rm ph})\\
&\times\frac{dP_{\rm bl}}{d\Gamma}d\Gamma dL_{\rm ph}.
\end{split}
\label{eq:all_neu_input}
\end{equation}
\end{linenomath*}
Meanwhile, using the LFs for luminosity-dependent density evolution (LDDE) models and parameters provided by \cite{ajello2012luminosity,ajello2013cosmic}, we successfully reproduced the redshift evolution of FSRQ and BL Lac luminosity densities illustrated in the Figure 6 of \cite{ajello2013cosmic}. At this stage, during the integration of $L_{\rm ph}$, we set the maximum and minimum luminosities to be $10^{50}\ \rm erg\ s^{-1}$ and $10^{40}\ \rm erg\ s^{-1}$, respectively. We also found that the results are consistent with the uncertainties in \cite{ajello2013cosmic} when the limits of the integration were varied by one or two orders of magnitude. Another thing that we need to keep in mind is that we assume the $Fermi$-LAT-unresolved blazars share the identical LFs with the resolved ones. \cite{2015ApJ...810...14A} pointed that the index distributions for different blazar classes both for the detected ones and undetected ones are slightly different: the photon spectra of newly-detected FSRQs are slightly softer than the 2LAC ones ($\Delta\Gamma<0.1$) while in contrast there is no significant spectral difference between the two sets of BL Lacs. For the completeness, we also consider a deviation, e.g., 0.2, of the photon spectral index from the best-fit values provided by \cite{ajello2015origin,ajello2012luminosity,ajello2013cosmic}. Such a test reveals that the resulting $\mathcal F(\gamma_{\rm lw})$ remains almost unchanged under a slight derivation of $\Gamma$.

Assuming the neutrino spectra from all blazars have the similar power-law form, e.g., $\varepsilon_\nu^2\Phi_{\varepsilon_\nu}\propto\varepsilon_\nu Q_{\varepsilon_\nu}^{({\rm bl,R/all})}\propto\varepsilon_\nu^{2-s}$, and using the comoving neutrino luminosities $\varepsilon_\nu Q_{\varepsilon_\nu}^{({\rm bl,all})}(z,{\gamma_{\rm lw}})$ and $\varepsilon_\nu Q_{\varepsilon_\nu}^{({\rm bl,R})}(z,{\gamma_{\rm lw}})$, the all-flavor neutrino fluxes from $Fermi$-LAT-resolved and all blazars at earth are expected to be 
\begin{linenomath*}
\begin{equation}
E_\nu^2\Phi_{E_\nu}^{(\rm bl,R/all)}({\gamma_{\rm lw}})=\frac{c}{4\pi}\int dz\ \frac{\varepsilon_\nu Q_{\varepsilon_\nu}^{({\rm bl,R/all})}(z,{\gamma_{\rm lw}})}{(1+z)}\left|\frac{dt}{dz}\right|,
\label{eq:neu_flux}
\end{equation}
\end{linenomath*}
where $\varepsilon_\nu=(1+z)E_\nu$. Hence, we can write down the fraction of $Fermi$-LAT-resolved blazars to the cumulative neutrino flux in a simple way that depends only on ${\gamma_{\rm lw}}$,
\begin{linenomath*}
\begin{equation} 
\mathcal F({\gamma_{\rm lw}})=\frac{E_\nu^2\Phi_{E_\nu}^{(\rm bl,R)}({\gamma_{\rm lw}})}{E_\nu^2\Phi_{E_\nu}^{(\rm bl,all)}({\gamma_{\rm lw}})}.
\label{eq:frac_resolved}
\end{equation}
\end{linenomath*}
\cite{ajello2015origin} presented the best-fit parameters in the blazar luminosity functions $\phi_{\rm bl}$, which enables us to compute $\mathcal F({\gamma_{\rm lw}})$. Since the redshift correction to the energies leads to one extra term $(1+z)^{2-s}$ to the integrand in equation \ref{eq:neu_flux} and another factor $(1+z)^{-1}$ to the integrated flux in equation \ref{eq:flux}, we conclude that, as a consequence, low-redshift blazars become more important when $s=2.5$, in comparison with the $s=2$ case. Therefore, considering nearby blazars are easier to be detected, a steeper neutrino spectrum predicts a larger $\mathcal F({\gamma_{\rm lw}})$, which is confirmed by the thin lines in Figure \ref{fig:fraction}. Moreover, noting that the selection of the minimum and maximum luminosities, e.g., $L_{\rm ph,\min}$ and $L_{\rm ph,\max}$ of a blazar is arbitrary, we tested the reliability of $\mathcal F(\gamma_{\rm lw})$ by varying the integral limits and found that} the results are not sensitive to $L_{\rm ph,\max}$ {  and $\mathcal F({\gamma_{\rm lw}})$ does not change dramatically} in the range ${\gamma_{\rm lw}}\lesssim1.0$ as $L_{\rm ph,min}$ increases from $10^{41}\ \rm erg\ s^{-1}$ to $10^{43}\ \rm erg\ s^{-1}$, as shown in Figure \ref{fig:fraction}.  Intuitively, a lower $L_{\rm ph,\min}$ implies that more low-luminosity blazars in the sample are less likely to be detected. Also, for a weaker luminosity dependance (${\gamma_{\rm lw}}\lesssim1.0$), the low-luminosity blazars dominate the luminosity density due to the large population. The combined effect is that $\mathcal F({\gamma_{\rm lw}})$ decreases in the range ${\gamma_{\rm lw}}\lesssim1.0$. Remarkably, from Figure \ref{fig:fraction}, we can conclude that the contribution from $Fermi$-LAT-resolved blazars is nearly the same as the neutrino flux from all blazars when ${\gamma_{\rm lw}}$ is larger than 1.0. The reason is that, assuming a stronger luminosity dependance ({  on other words, a higher $\gamma_{\rm lw}$}), the brighter blazars become increasingly important. These high-luminosity blazars have a higher chance to be detected and in this case the neutrinos luminosity densities from $Fermi$-LAT-resolved blazars and all blazars are comparable. 

To compute the upper limit of cumulative neutrino flux from all blazars, we use the existing constraints, $E_\nu^2\Phi_{E_\nu}^{(\rm 2LAC, stacking)}$ and $E_\nu^2\Phi_{E_\nu}^{(\rm 3LAC,stacking)}$, from blazar stacking analyses and point-source searches \citep{aartsen2017contribution,hooper2018active}, which are based on $Fermi$-LAT 2LAC and 3LAC blazars. Combining these existing limits with the fraction of the neutrino flux from $Fermi$-LAT-resolved blazars, we estimate the upper limits of all-blazar contributions from $Fermi$-LAT 2LAC and 3LAC analysis,
\begin{linenomath*}
\begin{equation}
E_\nu^2 \Phi_{E_\nu}^{(\rm2LAC/3LAC)}=\frac{E_\nu^2\Phi_{E_\nu}^{(\rm 2LAC/3LAC,stacking)}}{\mathcal F{({\gamma_{\rm lw}})}}.
\label{eq:all_blazar_limits}
\end{equation}
\end{linenomath*}
{The stacking results themselves have some model dependence. Here, to obtain conservative limits, we adopt the results based on the equal flux weighting for $E_\nu^2\Phi_{E_\nu}^{(\rm 2LAC/3LAC,stacking)}$. In general this gives conservative limits, and the luminosity weighting improves the constraints. We will see that, even in this most conservative case, the combined constraints of stacking and multiplet analysis are stringent.}

\begin{figure}
\includegraphics[width=0.5\textwidth]{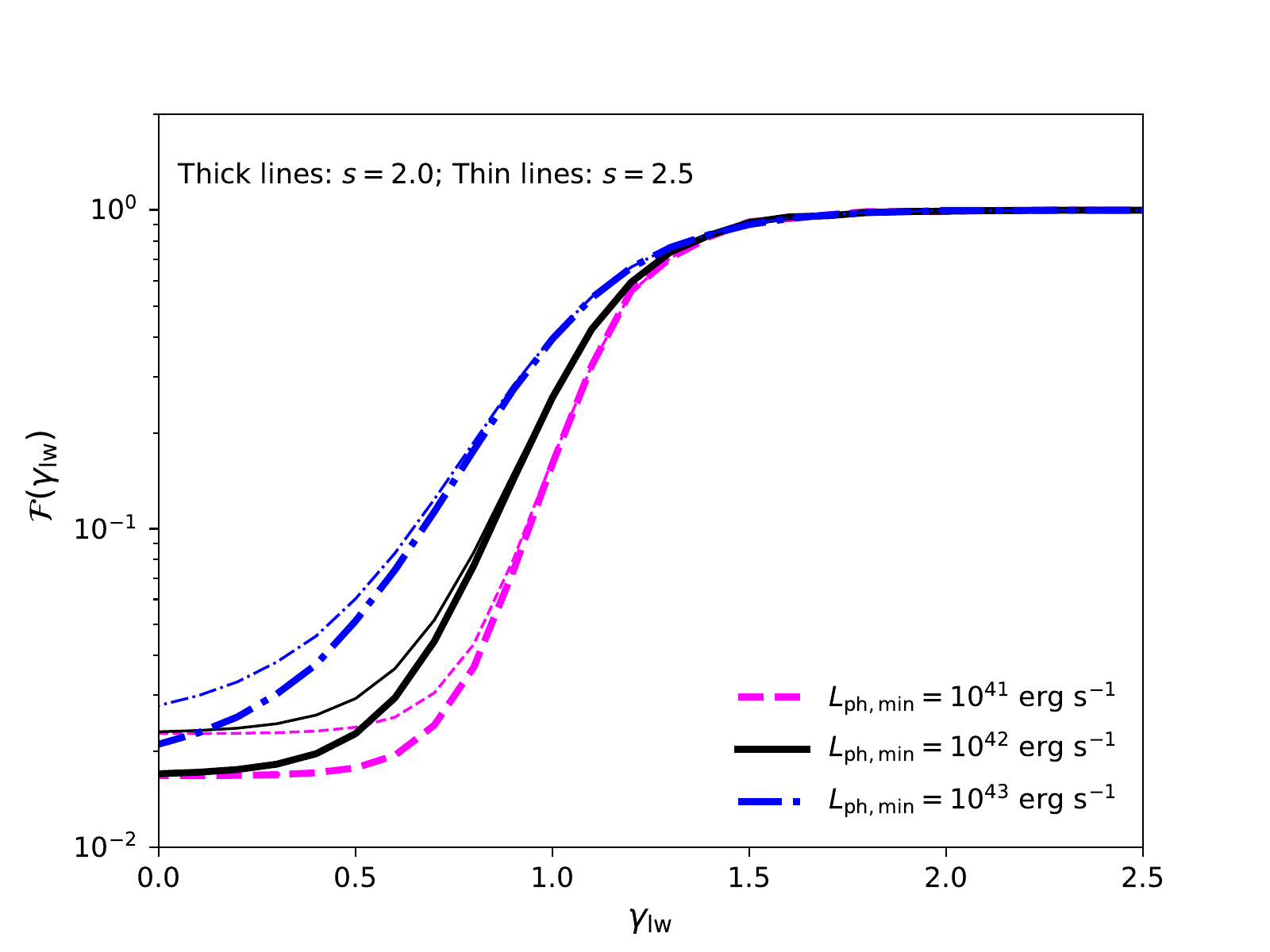}
\caption{The fraction of $Fermi$-LAT-resolved blazars in the cumulative neutrino flux, $\mathcal F({\gamma_{\rm lw}})$. The thick and thin lines are calculated for the neutrino spectral indices $s=2.0$ and $s=2.5$. The blue dashed, black solid and red dash-dotted lines correspond to the minimum luminosities $L_{\rm ph,\min}=10^{41}\ \rm erg\ s^{-1}$, $10^{42}\ \rm erg\ s^{-1}$ and $10^{43}\ \rm erg\ s^{-1}$, respectively. The upper limit is fixed to be $L_{\rm ph,max}=10^{50}\ \rm erg\ s^{-1}.$} \label{fig:fraction}
\end{figure}

\begin{figure*}
\includegraphics[width=0.5\textwidth]{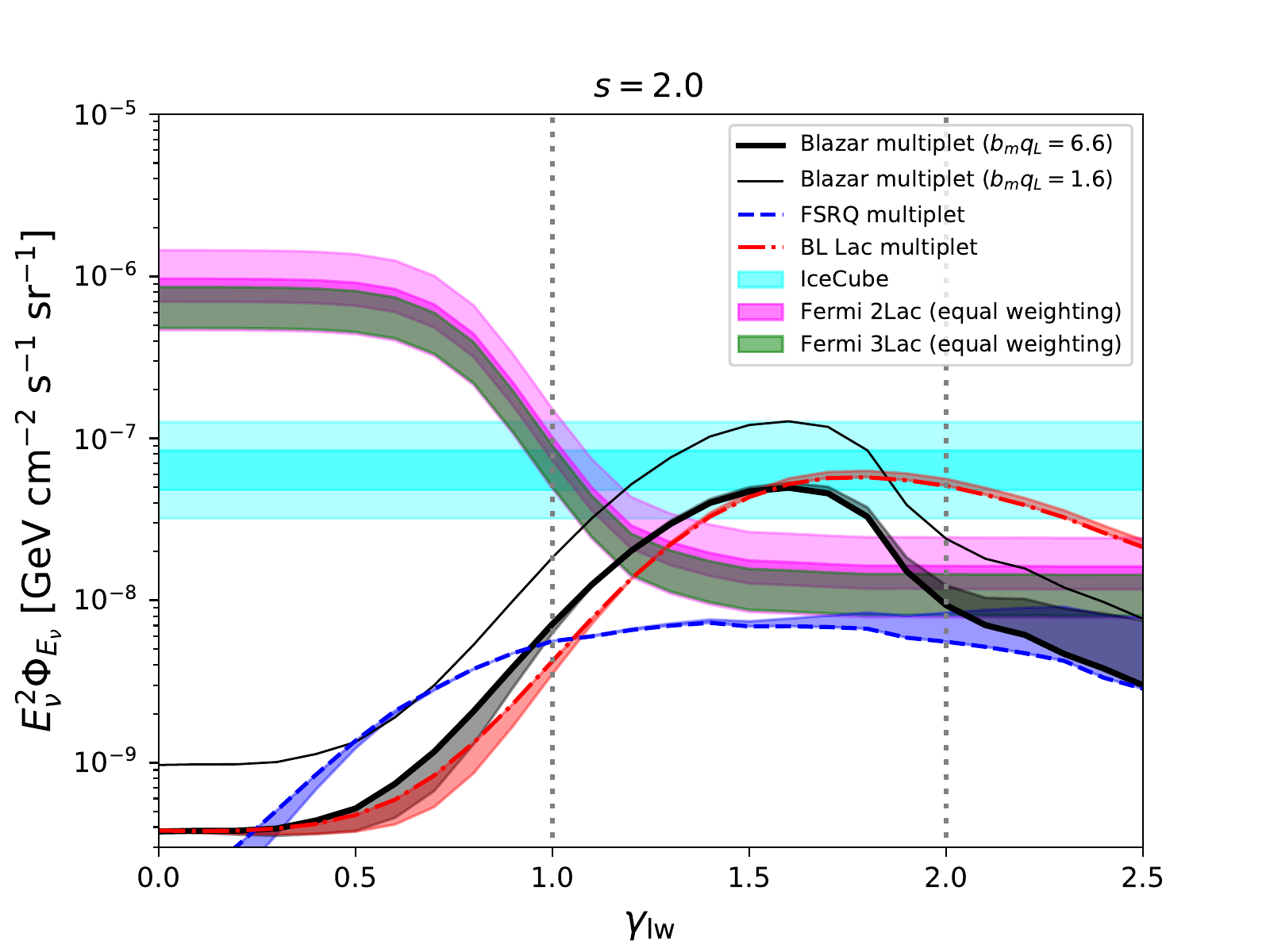}
\includegraphics[width=0.5\textwidth]{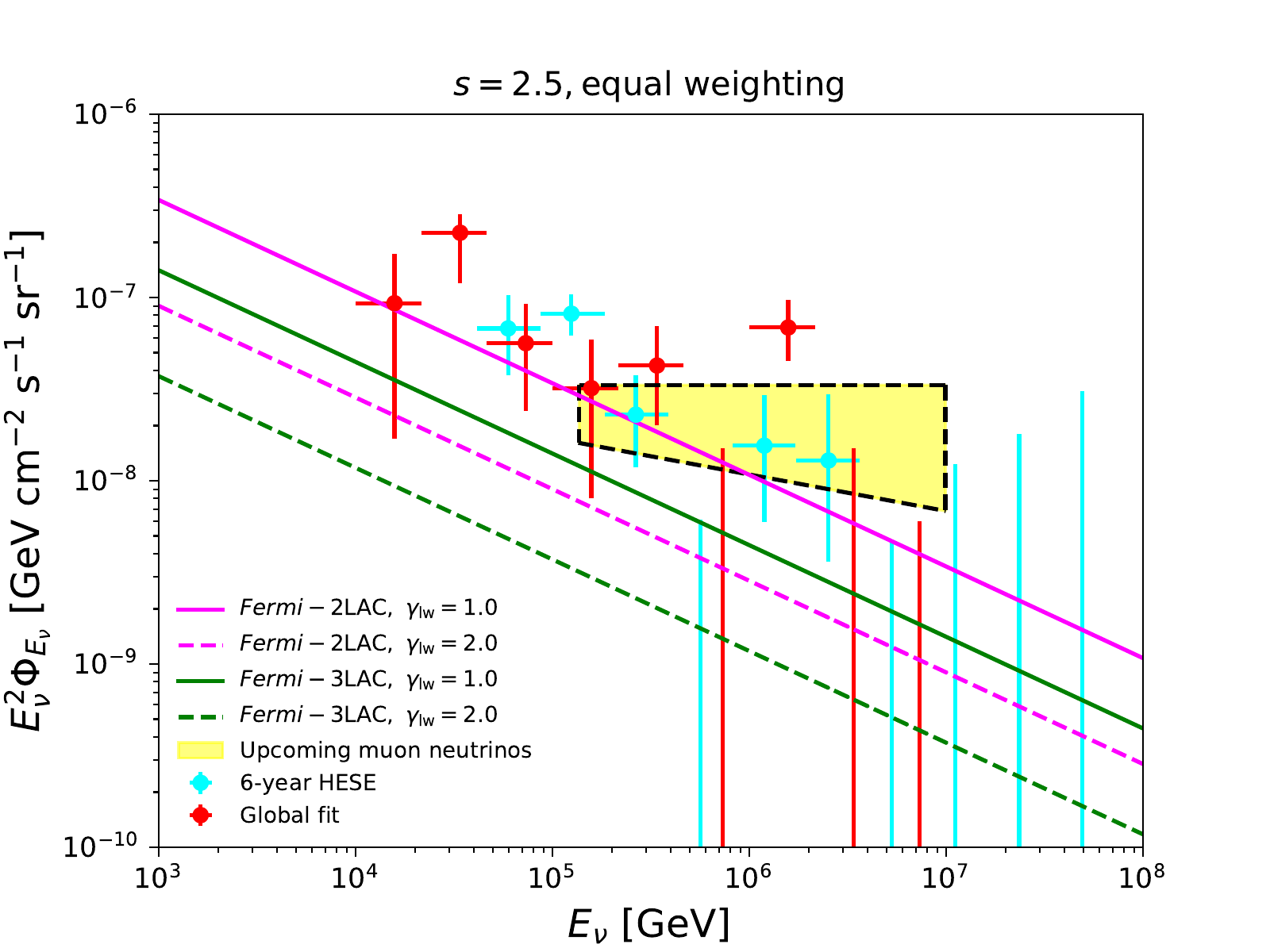}
\caption{All curves and data points in this figure illustrate all-flavor neutrino fluxes. Left panel: Stacking constraints on the contributions of all blazars to the cumulative neutrino flux ($L_{\rm ph,min}=10^{42}\rm\ erg\ s^{-1}$ is used) and high-energy neutrino multiplet constraints on the blazar contributions in the neutrino sky for an $\varepsilon_\nu^{-2}$ neutrino spectrum. The magenta and green areas correspond to the all-blazar upper limit from $Fermi$-LAT-2LAC and $Fermi-$3LAC equal weighting analysis, respectively. The cyan horizontal area shows the cumulative neutrino flux detected by IceCube. The blue dashed, red dash-dotted and thick black lines illustrate the $m\geq2$ multiplet constraints for FSRQs, BL Lacs and all blazars whereas the corresponding areas show the uncertainties. The thin black line is the $m\geq3$ multiplet constraint for all blazars. Right panel: the energy-dependent upper limits from the stacking analysis for the all-blazar contributions, assuming a neutrino spectral index $s=2.5$.
\label{fig:limits}}
\end{figure*}

Figure \ref{fig:limits} illustrates the upper limits for the all-blazar neutrino flux from $Fermi$-LAT 2LAC and $Fermi$-LAT 3LAC analysis. {We show all-flavor neutrino fluxes for all curves and data points in this figure.} In the left panel, we assume $s=2$ for the neutrino spectrum. In this case, the stacking analysis of $Fermi$-LAT-2LAC blazars gives $1.2\times10^{-8}\lesssim E_\nu^2\Phi_{E_\nu}^{(\rm 2LAC,stacking)}\lesssim1.6\times10^{-8}$ (in the unit of ${\ \rm GeV\ cm^{-2}\ s^{-1}\ sr^{-1}}$, hereafter). The corresponding upper limits for all blazars calculated using equation \ref{eq:all_blazar_limits} are illustrated as the magenta area. The green area in the left panel shows the constraints derived from $Fermi-$LAT 3LAC analysis which predicts $8.0\times10^{-9}\lesssim E_\nu^2\Phi_{ E_\nu}^{(\rm 3LAC,stacking)}\lesssim1.4\times10^{-8}$. For the illustration purpose, we include the IceCube all-flavor neutrino flux $4.8\times10^{-8}\lesssim E_\nu^2\Phi_{E_\nu}^{(\rm IC)}\lesssim8.4\times10^{-8}$ in Figure \ref{fig:limits} (the cyan area). {To avoid underestimating the upper limits due to the uncertainties of the existing results, we introduced a 50\% uncertainty to the constraints derived from stacking analysis, which broadens the areas in the left panel of figure \ref{fig:limits}.} The right panel shows the
energy-dependent upper limits for an $\varepsilon_{\nu}^{-2.5}$ neutrino spectrum. The solid lines are obtained by assuming ${\gamma_{\rm lw}}=1.0$ whereas the dashed lines correspond to the case ${\gamma_{\rm lw}}=2.0$. The upper limits from $Fermi$-LAT 2LAC and 3LAC analysis are illustrated as magenta lines and green lines, respectively. In this figure, we showed also the all-flavor neutrinos flux \citep[red points;][]{aartsen2015combined,aartsen2016observation}, the 6-year high-energy starting events \citep[cyan points;][]{Aartsen:2017mau} and the the best fit to the upcoming muon neutrinos scaled to three-flavor case (yellow area).
% where the neutrino fluxes from $Fermi-$2LAC and $Fermi-$3LAC analysis are constrained in the ranges $1.28\times10^{-8}\lesssim\varepsilon_\nu^2\Phi_{\varepsilon_\nu}^{(\rm 2LAC)}\lesssim1.79\times10^{-8}$ and $2.38\times10^{-9}\lesssim\varepsilon_\nu^2\Phi_{\varepsilon_\nu}^{(\rm 3LAC)}\lesssim7.4\times10^{-9}$, respectively. In these calculations, we assume $L_{\rm ph,min}=10^{42}\rm\ erg\ s^{-1}$ and $L_{\rm ph,max}=10^{50}\rm\ erg\ s^{-1}$. 
The previous discussion reveals that $\mathcal F({\gamma_{\rm lw}})$ may depend on $L_{\rm ph,min}$ moderately, when ${\gamma_{\rm lw}}$ is smaller than 1.0. We will further demonstrate in \S\ref{sec:multiplet} that, in the range of ${\gamma_{\rm lw}}\lesssim1.0$, the neutrino multiplet constraints are more stringent than the upper limits derived from the stacking analyses, which manifests its complementarity in constraining the cumulative neutrino flux from all blazars over a wide range of $\gamma_{\rm lw}$.

\section{Implications of High-Energy Neutrino Multiplet Limits}\label{sec:multiplet}
Here, we present another type of constraints on the origins of IceCube diffuse neutrinos, using the negative results from the clustering test of neutrino-induced muon track events. These high-energy track events are generally detected by IceCube with the angular resolution $\sim0.5\rm\ deg$, which enables us to determine the incoming directions and perform clustering analysis on their time and spatial distributions. So far, all the clustering tests based on high-energy muon neutrinos have found no statistically significant evidence of clustering in the arrival distribution of neutrinos \citep{aartsen2014observation,2015PhRvL.115h1102A,2017ApJ...835..151A,2019EPJC...79..234A}. 

In this section, we investigate the implications of the non-detection of neutrino multiplet sources, and consider the limits on blazar contributions to the cumulative neutrino background. To achieve this goal, we follow \cite{murase2016constraining} and write down the limits on the effective source densities. The formalism presented by \cite{murase2016constraining} is applicable to blazars with a general luminosity weighting $L_\nu\propto (L_{\rm ph})^{\gamma_{\rm lw}}$ since the functions $L_\nu(dN_{\rm bl}/dL_{\rm ph})\propto (L_{\rm ph})^{\gamma_{\rm lw}+1}\phi_{\rm bl}$ are sharply peaked around some effective luminosities $L_{\rm ph}^{\rm eff}$, which demonstrates that the effective source densities and the neutrino luminosity densities are well defined and constrained. Below, we define these crucial quantities and derive the neutrino multiplet constraints for our blazar case.

Assuming the number of sources that produce more than $k-1$ multiplet events is $N_{m\geq k}$, the constraint from the non-detection of $m\geq k$ multiplet events can be obtained by requiring $N_{m\geq k}\leq1$.  \cite{murase2016constraining} studied the implications to the neutrino sources using the absence of ``high-energy'' multiplet neutrino sources, and calculated the upper limit on the local source number density for an $\varepsilon_\nu^{-2}$ neutrino spectrum,
\begin{linenomath*}
\begin{equation}\begin{split}
n_0^{\rm eff}&\lesssim1.9\times10^{-10}{\rm\ Mpc^{-3}}\left(\frac{\varepsilon_\nu L_{\varepsilon_\nu}^{\rm ave}}{10^{44}\rm \ erg\ s^{-1}}\right)^{-3/2} \left(\frac{b_mq_L}{6.6}\right)^{-1}\\
&\times\left(\frac{F_{\rm lim}}{10^{-9.2}\rm \ GeV\ cm^{-2}\ s^{-1}}\right)^{3/2}\left(\frac{2\pi}{\Delta\Omega}\right),
\end{split}
\label{eq:n_eff_constraint}
\end{equation}
\end{linenomath*}
where $\varepsilon_\nu L_{\varepsilon_\nu}^{\rm ave}$ is the time-averaged neutrino luminosity of the source, $F_{\rm lim}\sim(5-6)\times10^{-10}{\rm \ GeV\ cm^{-2}\ s^{-1}}$ is the 8-year IceCube point-source sensitivity at the 90\% confidence level \citep{aartsen2017icecube}, $q_L\sim1-3$ denotes a luminosity-dependent correction factor, $\Delta \Omega$ represents the sky coverage of the detector and the details of $m\geq k$ {  neutrino multiplet constraints} are encoded in the factor $b_m$. \cite{murase2016constraining} find $b_m\simeq6.6$ for $m\geq2$ multiplets and $b_m\simeq1.6$ for triplets or higher multiplets (e.g., $m\geq3$). 
{Note that the point-source sensitivity enters the above expression but the numerical results are obtained by calculating the number of tracks using the muon effective area  \citep{murase2016constraining}. } 

\begin{figure}
\includegraphics[width=0.5\textwidth]{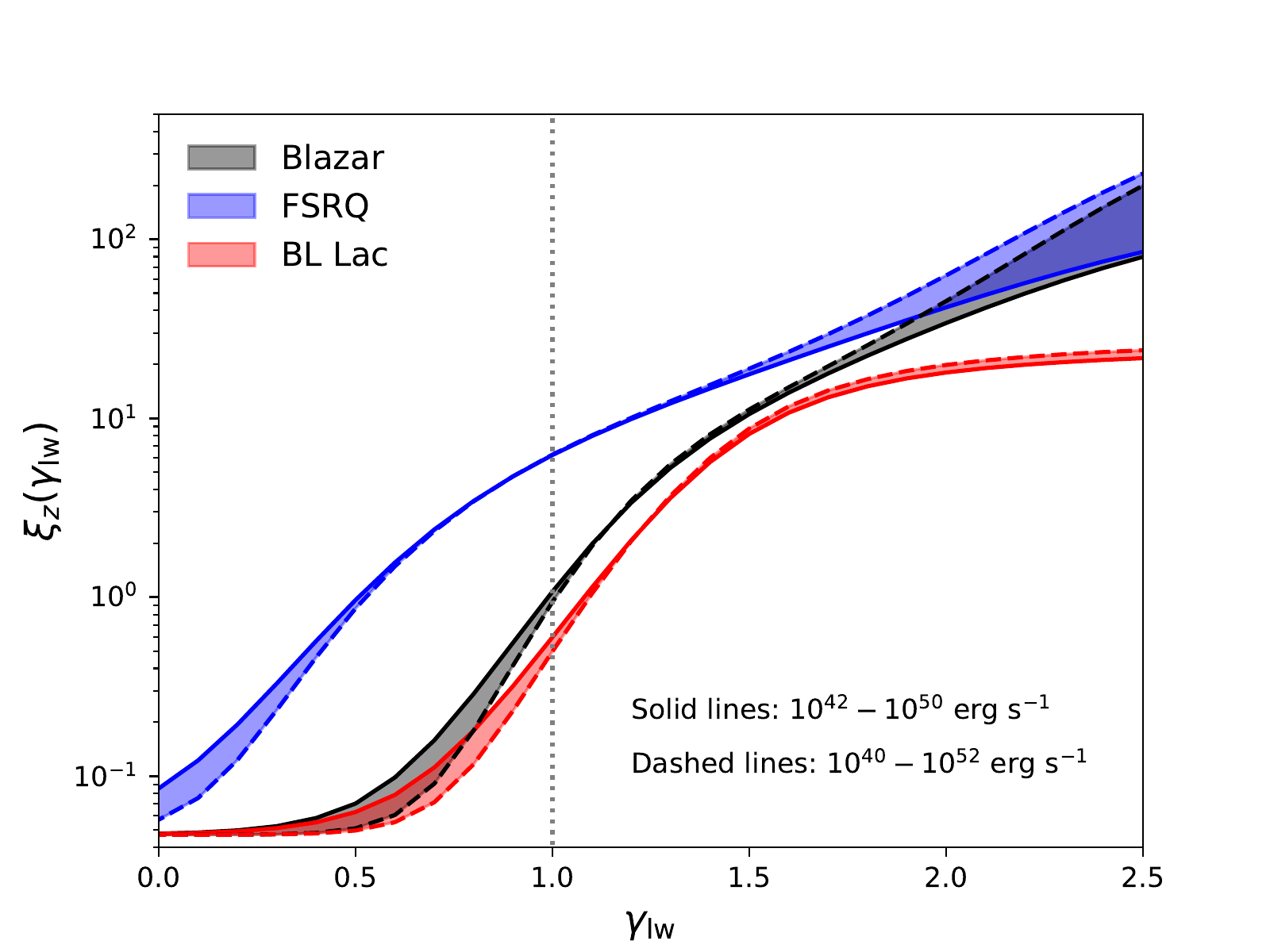}
\caption{The redshift evolution factor $\xi_z$ for FSRQs (blue area), BL Lacs (red area) and all blazars (black area). The solid and dashed boundaries correspond to different schemes of $L_{\rm ph,min}$ and $L_{\rm ph,max}$.}
\label{fig:xi}
\end{figure}

\begin{figure*}
\includegraphics[width=0.5\textwidth]{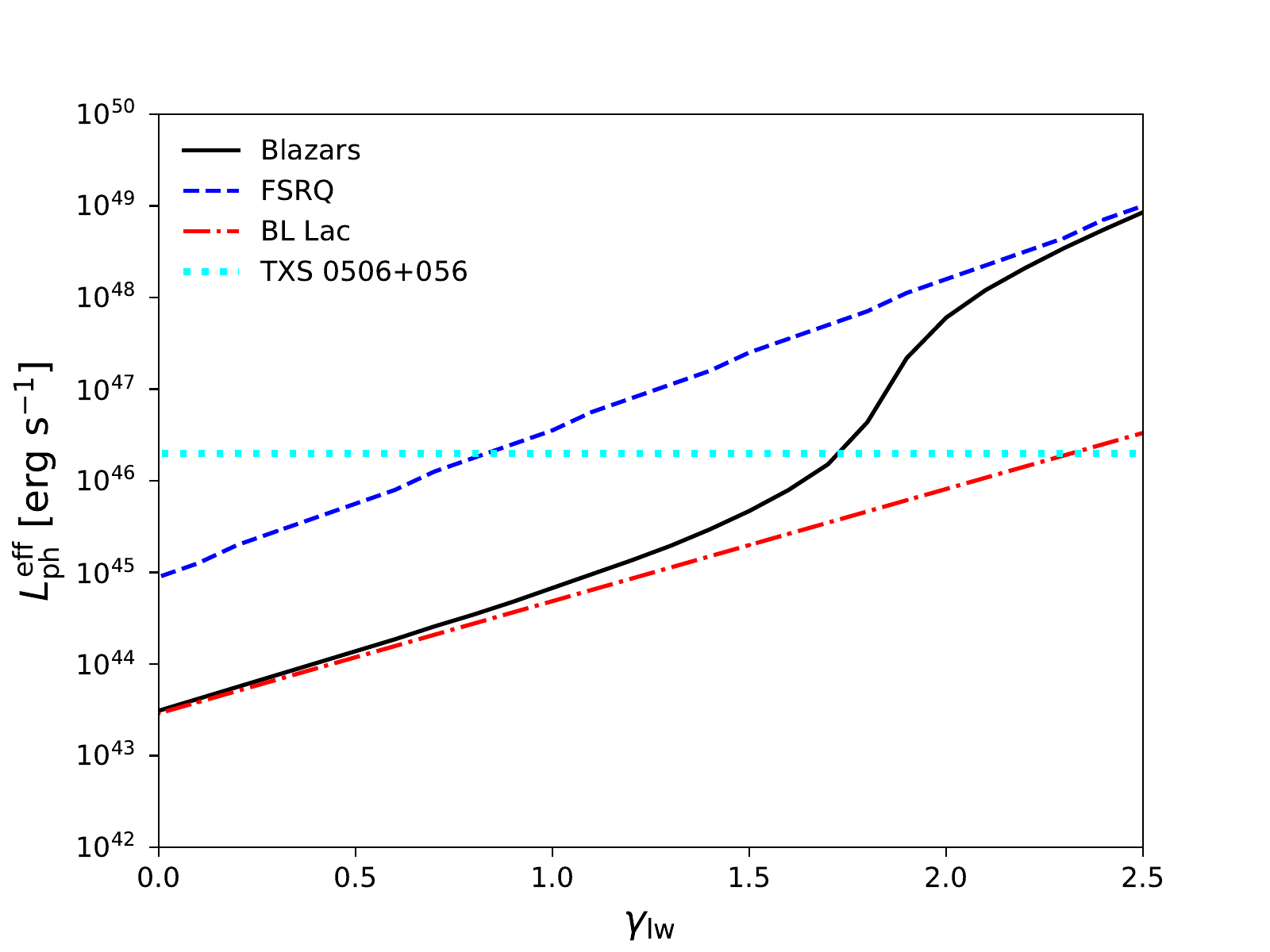}
\includegraphics[width=0.5\textwidth]{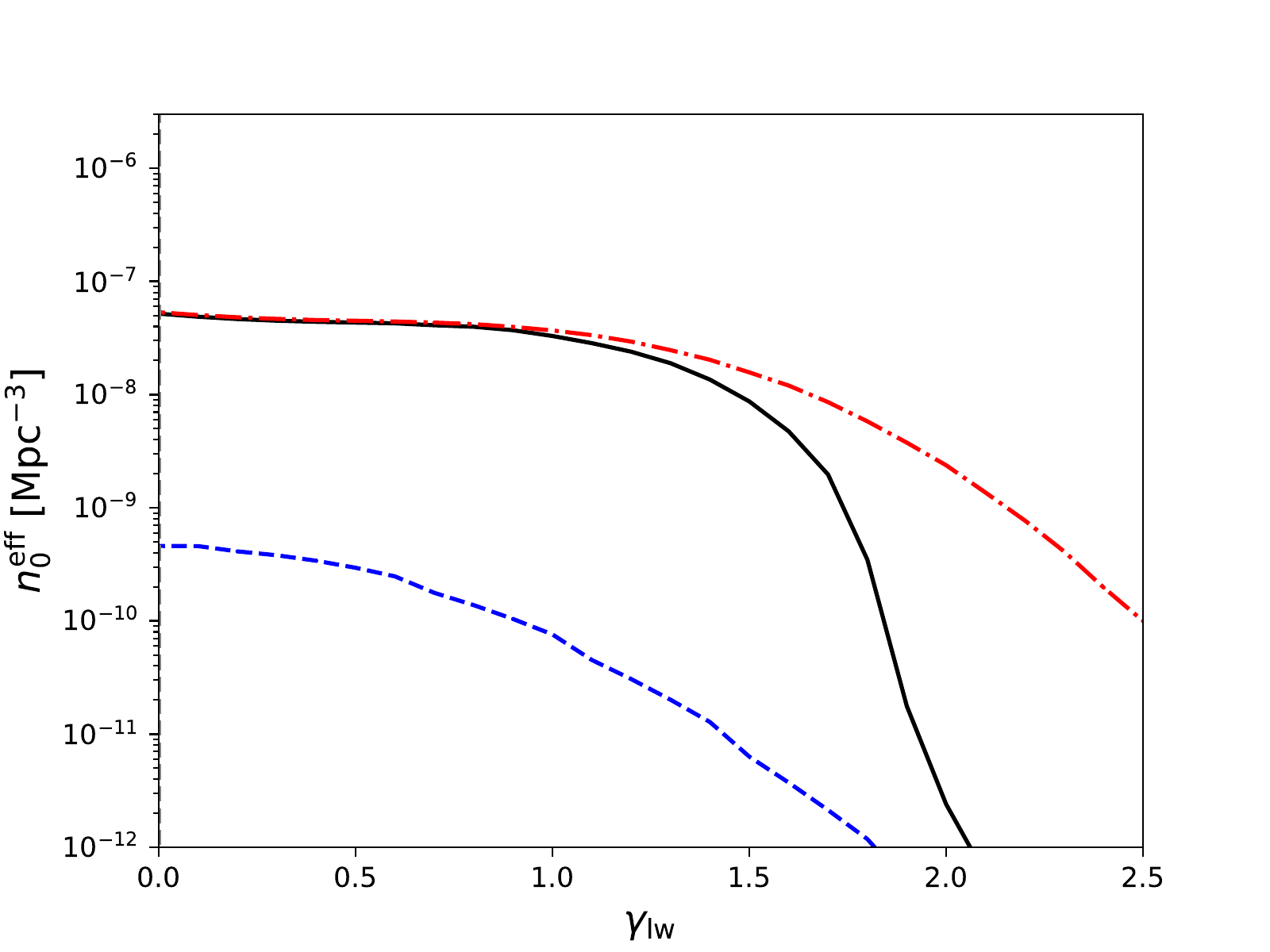}
\caption{Left panel: The effective gamma-ray luminosity for FSRQs (blue dashed line), BL Lacs (red dash-dotted line) and all blazars (black line). {The dotted horizontal line indicates the luminosity of TXS 0506+056, one blazar that features an intermediate luminosity \citep[$L_{\rm TXS}\simeq10^{46.3}\rm\ erg\ s^{-1}$; ][]{murase2018blazar}.} Right panel: The effective local number densities for different source classes. The line styles in this panel have the same meaning as the left panel.}
\label{fig:n_eff}
\end{figure*}

The purpose of this work is to explore the implications for blazar models using existing equations from previous work without making new analyses on multiplet sources. We simply use the results of the previous analysis by \cite{murase2016constraining}, which gives the upper limit on the effective number density, $n_{0}^{\rm eff}(\varepsilon_\nu L_{\varepsilon_\nu}^{\rm ave})$. 
Moreover, another reason that we choose this approach is that these results are also consistent with the latest limits on transient sources \citep[after the number density is converted into the rate density, e.g.,][]{2019PhRvL.122e1102A}~\footnote{The limit on the rate density of neutrino transients accounting for the diffuse flux is $\rho_{0}^{\rm eff}\gtrsim1.7\times10^{4}\,{\rm Gpc^{-3}}\,{\rm yr^{-1}}\,{(b_mq_L/6.6)}^2{({\Delta \Omega/2\pi)}^2{(T_{\rm obs}/8~{\rm yr})}^2}{(\xi_z/0.7)}^{-3}\phi^{-3}_{\rm lim, -1}$\\${\rm max}[N_{\rm fl},1]$, where $N_{\rm fl}\approx f_{\rm fl}T_{\rm obs}/t_{\rm dur}\approx T_{\rm obs}/\Delta T_{\rm fl}$ is the number of flaring periods and $\Delta T_{\rm fl}$ is the typical flare interval~\citep{murase2018blazar}. For $\Delta T_{\rm fl}\lesssim T_{\rm obs}$, the density and diffuse limits become similar to those for steady sources. Substituting the time-averaged sensitivity gives conservative results because of $F_{\rm lim}>\phi_{\rm lim}/T_{\rm obs}$. For $\Delta T_{\rm fl}\gtrsim T_{\rm obs}$, we expect $\rho_0^{\rm eff}T_{\rm obs}\approx n_{0}^{\rm eff}(T_{\rm obs}/\Delta T_{\rm fl})$. Because of $n_{0}^{\rm eff}(T_{\rm obs}/\Delta T_{\rm fl})\lesssim n_{0}^{\rm eff}$, the limits for steady sources can be regarded as conservative.}

One can write the limit on the cumulative neutrino flux from the sources as a function of $n_0^{\rm eff}$ and the redshift evolution factor $\xi_z$ \citep{murase2018blazar}:
\begin{linenomath*}
\begin{eqnarray}
%\begin{split}
E_\nu^2\Phi^{\rm (m)}_{E_\nu}&\approx& \frac{3\xi_zct_H}{4\pi}n_{0}^{\rm eff}(\varepsilon_\nu L_{\varepsilon_\nu}^{\rm ave})\nonumber\\
&\lesssim& 6.9 \times10^{-9}~{\rm GeV}{\rm cm}^{-2}{\rm s}^{-1} {\rm sr}^{-1}~\left(\frac{\Delta \Omega}{2\pi}\right)^{2/3}\left(\frac{\xi_z}{0.7}\right)\nonumber\\
&\times& \left(\frac{b_mq_L}{6.6}\right)^{-2/3} \left(\frac{n_0^{\rm eff}}{10^{-7}\ \rm Mpc^{-3}}\right)^{1/3}\nonumber\\
&\times&\left(\frac{F_{\rm lim}}{10^{-9.2}\rm \ GeV\ cm^{-2}\ s^{-1}}\right),
%\end{split}
\label{eq:flux_constraint}
\end{eqnarray}
\end{linenomath*}
where $t_H$ is the Hubble time. In this expression, $\xi_z$ represents the redshift weighting of the neutrino luminosity of the sources and can be evaluated through~\citep{waxman1998high} 
\begin{linenomath*}
\begin{equation}
\xi_z({\gamma_{\rm lw}})=\frac{\int dz (1+z)^{-1}\left|\frac{dt}{dz}\right|f(z,{\gamma_{\rm lw}})}{\int dz \left|\frac{dt}{dz}\right|},
\label{eq:xi_z}
\end{equation}
\end{linenomath*}
where $f(z,{\gamma_{\rm lw}})$ is the redshift evolution function of the neutrino luminosity density normalized to unity at $z=0$ for the luminosity correlation $L_\nu\propto (L_{\rm ph})^{{\gamma_{\rm lw}}}$, e.g., for blazars we have $f^{\rm (bl)}(z,{\gamma_{\rm lw}})=[\varepsilon_\nu Q_{\varepsilon_\nu}^{(\rm bl,all)}(z,{\gamma_{\rm lw}})]/[\varepsilon_\nu Q_{\varepsilon_\nu}^{(\rm bl,all)}(0,{\gamma_{\rm lw}})]$. 
Similarly, we can also calculate the $\xi_z$ for the blazar subclasses, FSRQs and BL Lacs using the luminosity functions from \cite{ajello2012luminosity,ajello2013cosmic}. The black, blue and red areas in Figure \ref{fig:xi} illustrate the redshift evolution factor $\xi_z({\gamma_{\rm lw}})$ for all blazars, FSRQs and BL Lacs, respectively. When ${\gamma_{\rm lw}}=1$, we find $\xi_z\sim7-8$ for the gamma-ray luminosity density evolution of FSRQs and $\xi_z\sim0.6-0.7$ for that of BL Lacs, which are consistent with the values found by \cite{murase2014diffuse} and \cite{murase2016constraining}. 
The solid and dashed boundaries in Figure \ref{fig:xi} correspond to the sample schemes, $(L_{\rm ph,min}=10^{42}{\ \rm erg\ s^{-1}},\ L_{\rm ph,max}=10^{50}\ {\rm erg\ s^{-1}})$ and $(L_{\rm ph,min}=10^{40}{\ \rm erg\ s^{-1}},\ L_{\rm ph,max}=10^{52}\ {\rm erg\ s^{-1}})$, respectively. If ${\gamma_{\rm lw}}$ is lower than 1.0, low-luminosity sources at lower redshift contribute a significant component to the total neutrino luminosity density, therefore, a smaller $L_{\rm ph,\min}$  results in a smaller $\xi_z$. On the contrary, a strong luminosity correlation with ${\gamma_{\rm lw}}\gtrsim1.5$ boosts the contribution from high-redshift bright blazars, which leads to a larger $f(z,{\gamma_{\rm lw}})$ at higher redshift and as a result makes $\xi_z$ larger, as $L_{\rm ph,max}$ increases. 

Besides the factor $\xi_z$, it is also necessary to calculate the effective local number density $n_0^{\rm eff}$, which characterizes the the number density of sources that dominate the neutrino luminosity density for each specified source population. In this work, we use the luminosity functions in combination with the luminosity weighting relation $L_{\nu}\propto (L_{\rm ph})^{{\gamma_{\rm lw}}}$ to estimate the effective number densities $n_0^{\rm eff}$ for blazars, FSRQs and BL Lacs. Here, we follow the procedure presented by \cite{murase2016constraining}. For each class of neutrino sources, we define an effective neutrino luminosity $L_\nu^{\rm eff}\propto (L_{\rm ph}^{\rm eff})^{\gamma_{\rm lw}}$ using the corresponding effective gamma-ray luminosity $L_{\rm ph}^{\rm eff}$ obtained by maximizing $(L_{\rm ph})^{{\gamma_{\rm lw}}}(dN/d{\ln L_{\rm ph}})=(L_{\rm ph})^{{\gamma_{\rm lw}}+1}\phi(L_{\rm ph},z=0)$, where $\phi(L_{\rm ph},z=0)$ is the local luminosity function of the sources that we are interested in. 
Since the function $(L_{\rm ph})^{\gamma_{\rm lw}+1}\phi(L_{\rm ph},z=0)$ has a maximum around its extreme point for each source population, we may regard blazars, FSRQs and BL Lacs as ``quasi-standard candle" sources, among which the neutrino productions are dominated by the sources distributed closely around one certain effective luminosity $L_{\rm ph}^{\rm eff}$. In this case, we have justified the applicability of the equation appeared in this section to constrain the neutrino fluxes from blazars and the subclasses. 
The left panel of Figure \ref{fig:n_eff} shows the effective gamma-ray luminosity densities for all blazars (black solid line), FSRQs (blue dashed line) and BL Lacs (red dash-dotted line). Intuitively, $L_{\rm ph}^{\rm eff}$ of FSRQ should be larger than that of BL Lacs since FSRQs are more luminous than BL Lacs. Moreover, the function $(L_{\rm ph})^{{\gamma_{\rm lw}}+1}\phi(L_{\rm ph},z=0)$ achieves its maximum at higher luminosity as ${\gamma_{\rm lw}}$ increases, which naturally explains the monotonic increase of $L_{\rm ph}^{\rm eff}({\gamma_{\rm lw}})$. Considering that low-luminosity BL Lacs dominate the neutrino luminosity density if the luminosity correlation is weak (e.g., ${\gamma_{\rm lw}}\lesssim 1$) whereas bright FSRQs become increasingly important as ${\gamma_{\rm lw}}$ increases, the blazar effective luminosity $L_{\rm ph}^{\rm eff}$ converges to the BL Lac case when ${\gamma_{\rm lw}}$ is less than 1.0 and then gradually approaches to the FSRQ curve, as is confirmed in Figure \ref{fig:n_eff}. With the effective neutrino/gamma-ray luminosity, we can write down the effective local number density of the sources
\begin{linenomath*}
\begin{equation}
n_{0}^{\rm eff}=\frac{1}{L_{\nu}^{\rm eff}}\int dL_{\rm ph}L_\nu(L_{\rm ph})\phi(L_{\rm ph},0).
\label{eq:n_eff}
\end{equation}
\end{linenomath*}
The right panel of Figure \ref{fig:n_eff} shows the effective number densities of all blazars (black solid line), FSRQs (blue dashed line) and BL Lacs (red dash-dotted line). As expected, BL Lacs dominate the number density and the blazar effective number density converges to BL Lac and FSRQ curves respectively {when} ${\gamma_{\rm lw}}\lesssim1.0$ and ${\gamma_{\rm lw}}\gtrsim2.0$. Different from $\mathcal F({\gamma_{\rm lw}})$ and $\xi_{z}$, $L_{\rm ph}^{\rm eff}$ and $n_{0}^{\rm eff}$ does not depend sensitively on the value of $L_{\rm ph,min}$ and $L_{\rm ph,max}$ in the range $0\lesssim{\gamma_{\rm lw}}\lesssim2.5$. To interpret this, we need to keep in mind that the former two quantities are determined by the integrations over $L_{\rm ph}$, {while $L_{\rm ph}^{\rm eff}$ depends} only on the shape/slope of the function $(L_{\rm ph})^{{\gamma_{\rm lw}}+1}\phi(L_{\rm ph},z=0)$. From the left panel of Figure \ref{fig:n_eff}, we find that $L_{\rm ph}^{\rm eff}$ lies roughly in the range $10^{43}-10^{49}\rm \ erg\ s^{-1}$ which is covered by the interval $10^{42}-10^{50}{\rm \ erg\ s^{-1}}$, {  the fiducial range used in our calculation}. Meanwhile, the integrand in equation \ref{eq:n_eff} peaks around $L_{\rm ph}^{\rm eff}$, therefore once the peak is included, the effective number density $n_0^{\rm eff}$ will not vary too much as the lower and upper bounds of the integral changes.

The above calculations provide the preliminary work and the ingredients needed for calculating the neutrino multiplet limits. Selecting $b_mq_L\simeq6.6$ for $m\geq2$ multiplets and $F_{\rm lim}\simeq10^{9.2}\ \rm GeV\ cm^{-2}\ s^{-1}$ for an $\varepsilon_\nu^{-2}$ neutrino spectrum, the blue dashed, red dashed-dotted and thick black lines in the left panel of Figure \ref{fig:limits} illustrate the neutrino multiplet limits for FSRQs, BL Lacs and all blazars, respectively. The blue, red and black areas shows the corresponding uncertainties due to $L_{\rm ph,min}$ and $L_{\rm ph,max}$, as discussed before. From this figure we find that the all-blazar multiplet constraint converges to the FSRQ case at higher ${\gamma_{\rm lw}}$ and to the BL Lac case if ${\gamma_{\rm lw}}$ is less than 1.0, just as expected. We also considered the upper limits for triplet or higher multiplets ($m\geq3$) by changing the value of $b_mq_L $ to $1.6$. In this case, the constraints {relax} to the thin black line. This consequence can be interpreted as the concession of allowing blazars to produce $m=2$ multiplet events. So far, all calculations on the multiplet constraints {were} based on the $\varepsilon_\nu^{-2}$ neutrino spectrum, and to extend the results to a general spectrum, e.g., $s=2.5$, detailed calculations on $F_{\rm lim}$ and $n_{0}^{\rm eff}$ (equation \ref{eq:n_eff}) are needed, and our results are conservative in this point. Therefore, in the right panel of
Figure \ref{fig:limits}, only upper limits inferred from stacking analysis are shown.

\section{Discussion}\label{sec:discussion}
In this paper, we considered how two types of analyses, namely stacking and multiplets, constrain on the contribution of blazars to the cumulative neutrino flux, assuming a generalized luminosity weighting $L_{\nu}\propto (L_{\rm ph})^{{\gamma_{\rm lw}}}$. Using the gamma-ray luminosity functions for blazars, FSRQs and BL Lacs, we estimated the ratio of the neutrino fluxes from $Fermi$-LAT-resolved blazars and from all blazars (including unresolved ones), $\mathcal F({\gamma_{\rm lw}})$, and the effective number densities, $n_0^{\rm eff}({\gamma_{\rm lw}})$, and the redshift evolution factor, $\xi_z$, for different source classes. The joint use of a stacking and multiplet analysis, as well as the use of a generalized luminosity function and inclusion of the effect of unresolved blazars, are new aspects which distinguish this analysis from previous ones.
The main results are summarized in Figure~\ref{fig:limits}. From this figure we found that the multiplet constraints are the most important at lower values of ${\gamma_{\rm lw}}$, e.g. ${\gamma_{\rm lw}}\lesssim1.0$, whereas all-blazar constraints derived from {  the existing stacking upper limits are more stringent for a stronger luminosity correlation, e.g., $\gamma_{\rm lw}\gtrsim1.5$}. The joint consideration of these two kinds of limits supports the extended argument that all blazars, including $Fermi-$unresolved ones, are unlikely to dominate the cumulative neutrino background for a generic correlation between the neutrino and gamma-ray luminosities, $L_{\nu}\propto (L_{\rm ph})^{\gamma_{\rm lw}}$, with the index $0\lesssim\gamma_{\rm lw}\lesssim2.5$. Canonical blazar models, which are physically motivated and based on the leptonic scenario, predict ${\gamma_{\rm lw}}\sim1.5-2.0$~\citep{murase2014diffuse}. Our results suggest that the stacking constraints are the most stringent for such physically motivated cases. 
The multiplet and stacking limits are ``complementary'', in the sense that these methods have their own advantages in different regimes, and in combination they provide a stronger and tighter constraint than previously, over a wide range of $\gamma_{\rm lw}$, as pointed out by \cite{murase2018blazar}. We also found that while the multiplet constraints are weaker at larger values of ${\gamma_{\rm lw}}$ they become more stringent again for ${\gamma_{\rm lw}}\gtrsim1.5$ due to the rapid decrease of the effective source density.

In this work, we focus on power-law spectra. The limits are stringent for the neutrino flux in the 0.1~PeV range and become weaker at higher energies. For example, neutrino multiplet limits are weaker if one is interested in the origin of $\sim1$~PeV neutrinos~\citep{murase2016constraining,murase2018blazar,palladino2019interpretation}. 
It is possible for blazars to explain the dominant fraction of PeV neutrinos by introducing a lower-energy cutoff of the proton maximum energy~\citep{2014JHEAp...3...29D}, although neutrinos at 0.1~PeV and lower energies should come from another population of the sources~\citep[e.g.,][]{Murase:2019vdl}.  

One of the uncertainties in this work comes primarily from the selection of the lower and upper limits of the luminosity integral, $L_{\rm ph,min}$ and $L_{\rm ph,max}$. As discussed above, we showed that these uncertainties are well controlled, and the final results are reliable if $L_{\rm ph,min}$ and $L_{\rm ph,max}$ are selected in the fiducial ranges $10^{40}-10^{42}\ \rm erg\ s^{-1}$ and $10^{50}-10^{52}\ \rm erg\ s^{-1}$, respectively. From the joint constraints illustrated in Figure \ref{fig:limits}, we conclude that blazars are disfavored {as a dominant source of} the cumulative neutrino flux measured by IceCube for a luminosity weighting $L_{\nu}\propto (L_{\rm ph})^{{\gamma_{\rm lw}}}$ with $0.0\lesssim{\gamma_{\rm lw}}\lesssim2.5$. Since different blazar models considered for explaining the cumulative neutrino flux can be commonly characterized by the correlation index ${\gamma_{\rm lw}}$ within this range, our calculations on the upper limits and effective number densities would provide rather general constraints for future studies of blazar neutrinos.

%%%%%%%%%%%%%%%%%%%%%%%%%%%%%%%%%%%%%%%%%%%%%%%%%%
%%%%%%%%%%%%%%%%%%%%%%%%%%%%%%%%%%%%%%%%%%%%%%%%%%

\begin{acknowledgements}
We thank Marco Ajello for useful discussion on the usage of the luminosity function and Nick Rodd for the useful communication. 
The work of K.M. is supported by the Alfred P. Sloan Foundation and NSF grants No. PHY-1620777 and No. AST-1908689, while that of C.C.Y and P.M. is supported by the Eberly Foundation.
\end{acknowledgements}

\bibliographystyle{aasjournal}
%\bibliographystyle{apj_8}
%input{msp.bbl}
\bibliography{references}

\end{document}